\documentclass [psfig,11pt]{article}
\usepackage{natbib,amsfonts,epsfig,amssymb,mathrsfs}
\usepackage{times}
\usepackage[margin=1.0in]{geometry}
\usepackage{hyperref}
\usepackage{color}
\usepackage[utf8]{inputenc}

\definecolor{red}{rgb}{.7,0,0}
\definecolor{blue}{rgb}{0,0,1}

\title{What's In a Survey?  Simulation-Induced Selection Effects in Astronomy}
\author{Sarah C. Gallagher and Chris Smeenk}
\date{\today}

\begin{document}

\abstract{Observational astronomy is plagued with selection effects that must be taken into account when interpreting data from astronomical surveys.  Because of the physical limitations of observing time and instrument sensitivity, datasets are rarely complete. However, determining specifically what is missing from any sample is not always straightforward.  For example, there are always more faint objects (such as galaxies) than bright ones in any brightness-limited sample, but faint objects may not be of the same kind as bright ones.  Assuming they are can lead to mischaracterizing the population of objects near the boundary of what can be detected.  Similarly, starting with nearby objects that can be well observed and assuming that objects much farther away (and sampled from a younger universe) are of the same kind can lead us astray. Demographic models of galaxy populations can be used as inputs to observing system simulations to create ``mock'' catalogues that can be used to characterize and account for multiple, interacting selection effects.  The use of simulations for this purpose is common practice in astronomy, and blurs the line between observations and simulations; the observational data cannot be interpreted independent of the simulations.  We will describe this methodology and argue that astrophysicists have developed effective ways to establish the reliability of simulation-dependent observational programs.  The reliability depends on how well the physical and demographic properties of the simulated population can be constrained through independent observations.  We also identify a new challenge raised by the use of simulations, which we call the ``problem of uncomputed alternatives.'' Sometimes the simulations themselves create unintended selection effects when the limits of what can be simulated lead astronomers to only consider a limited space of alternative proposals.
}
\maketitle

\section{Introduction}

Scientists have increasingly come to rely on computer simulations as an essential component of empirical research.  Philosophers have studied the epistemological role simulations play in a handful of fields, including high-energy physics and climate science.  They have discovered that recent research in both areas has blurred the boundaries between measurement, observation, experiment, and simulation.  This prompts a general question:  What are the risks associated with treating not just experience, but experience enhanced through simulations, as our primary epistemic authority and guide?

Philosophers have approached this general question by giving detailed assessments of the use of simulations in different domains. \citet{parker2020computer} considers the practice of data assimilation in climate science, in which empirical measurements are combined with simulations to generate a more complete characterization of the state of the atmosphere.  Based in part on a liberal account of what constitutes a measurement, Parker defends treating a description of the atmospheric state constructed in this fashion as a ``measurement'' even though it incorporates simulation outputs.  Several recent studies of the use of simulations in high-energy physics describe their essential role in designing and interpreting experiments.  Discoveries such as that of the Higgs boson at the Large Hadron Collider rely on intricate simulations of both the events that occur in the beam pipe, and how the decay products produced by these events interact with detectors.  These simulations are needed to characterize the background against which a novel signal can be detected and to select appropriate candidate events from the detectors.  Philosophers have debated the precise contributions of simulations, such as the extent to which the Higgs discovery logically or causally depends on them \citep{morrison2015,massimibhimji2015,bogetrust}. But it is not controversial that these cases illustrate the thorough integration of simulation into experiment and observation.  

We also take these studies to show that, at least in some cases, scientists have overcome simulation-dependence to achieve reliable results.  But exactly how reliability can be established depends on what role the simulations play in research. In Parker's case study, for example, the simulation has to generate a description of the atmospheric state that is sufficiently close to the unknown true atmospheric state for the relevant purposes.  The main challenge to establishing the reliability of using the simulated state is that of calibration.  To play a role similar to that of measurements, the simulation needs to provide not just an estimate of the state but also of the associated uncertainties. By contrast with instrumental results, however, atmospheric scientists generally do not have well-motivated uncertainty estimates for the simulated states \citep[][\S 7]{parker2020computer}.  Assessing how simulations contributed to the Higgs discovery, and their reliability, involves a quite different set of issues.  To find out whether we can preserve reliability while integrating simulations into observations, we need to first clarify what role simulations actually play in a given field.  Our aim below is to highlight and assess a distinctive role simulations play in astrophysics. 

Astrophysicists have used simulations to treat selection effects, which often arise in scientific fields that rely primarily on passive observations.  What inferences we can draw from an observed sample depend on whether it is a fair sample from the overall population, in the relevant respects.  A selection effect refers to any bias introduced by our methods of modeling and observing the population.  For example, pollsters who contact participants by phone have to determine whether people with phones answer in the same way as those without. Modeling and accounting for selection effects requires detailed background knowledge about the target system as well as the observational program, in order to assess and control for biases.  This rapidly becomes quite complex, and has led astrophysicists to develop sophisticated modeling techniques that employ simulations in two distinctive ways.  First, the (hypothetical) demographics of the target population of objects needs to be specified.  Sometimes these can be treated as the output of a simulation.  For example, several large-scale structure simulations evolve forward from an initial state in the early universe (constrained by observations) to yield a distribution of galaxies and other structures at later times. In most cases, however, the simulations do not yield sufficient information about the relevant target population, and so specifying the demographics involves, by necessity, further modeling, physically motivated extrapolations, or inspired guesswork. A second type of simulations model the integrated effects of the telescope, instruments, and observing program design. We will call these observing system simulations, and they model what we should expect to see given our assumption about the target population.  We characterize the cumulative impact of uncertainties as a ``selection effect,'' because failures in either type of simulation lead us astray in treating the actual observations as a fair sample.

As an example, mock ``true'' galaxy populations, informed by the outputs from physical simulations of galaxy populations evolving over time, can be ``observed'' using the known properties of an observational survey (specifying details regarding, e.g., detector sensitivities and criteria used to select target objects).  The actual output of the observational survey will then be compared to the simulated observations of the ``true'' population.   The model, ``true'' galaxy populations can be modified until the actual and simulated galaxy catalogues converge.  This methodology has enabled an efficient and sophisticated treatment that accounts for cumulative selection effects, as we will describe in more detail in \S2 below.  This practice leads to simulations being woven into the fabric of galaxy surveys and various other observational programs in both experimental design and data interpretation, and raises questions about the reliability of the results.  This role for simulations differs from their use to provide, for example, detailed models of specific types of astrophysical systems, and raises different challenges to assessing reliability.  Our argument complements recent work in philosophy that emphasizes the essential role of simulations in astrophysics \citep{anderl2018simplicity,jacquart2020observations}, albeit in different senses, and critically assesses how their reliability can be established \citep{gueguen2020robustness}.

One novel challenge to reliability is apparent in a different example: evolutionary simulations of mergers of galaxy pairs.  These simulations can be stopped and compared to real images to search for ``how plausibly'' explanations for how a particular observed structure could have formed.  In this use of simulations, as with their use in galaxy surveys, the scope of possibilities considered is often informed by observations that are themselves plagued with selection effects.  This is a familiar problem, even though it is difficult to account for all potential sources of systematic bias. But there is a second more subtle kind of selection effect that we will emphasize, that arises due to computational constraints.  If it is only possible to model a suite of mergers of two galaxies, the possibility that an observed system results from interactions among three or more galaxies may not even be explicitly acknowledged or considered.  This computational selection effect limits the space of hypotheses being considered and therefore influences the type of observing programs undertaken.  We will consider the ramifications of this kind of limitation, based on a detailed case study, in \S3, before stating our conclusions in the final section.  

\section{Selection effects in astrophysics}

Astronomers often count things.  This is typically the first step in the observational study of different types of objects, leading to quantitative measures of a population. This might include organizing objects (e.g., stars) into bins based on, for example, their intrinsic brightnesses.  But there are obstacles to getting an accurate count. It is almost always the case that luminous objects -- such as the most massive galaxies and the hottest main sequence stars -- are rare.  So, in a circular section of sky (which represents a cone in volume), the numbers for the most luminous objects are small and have correspondingly large uncertainties from counting statistics.  Intrinsically faint objects, such as low mass stars, will be much more numerous, but can only be probed to a much smaller volume before reaching the brightness limit of the survey.  If one did not take this observational selection effect into account and correct for the different volumes that are visible for objects of different luminosities, then one would get a very skewed understanding of the true distribution of objects of different luminosities.  

This concern was recognized early by \citet{Malmquist1922}, and is a well-known example of a selection effect.  If distances to objects are known (and thus observed brightnesses corrected to true luminosities), then Malmquist bias can be accounted for with a simple geometric correction for the relative volume at which one could detect objects of a given true luminosity.

The consequences of Malmquist bias are complicated when the numbers of intrinsically fainter objects are greater than the numbers of brighter objects (as is typical), and measurement uncertainties are taken into account.  If measurement uncertainties are symmetric about the measured value, and greater numbers exist at fainter fluxes, then objects will preferentially be scattered from fainter into brighter luminosity bins.  This effect, known as Eddington bias \citep{Eddington1913}, can also skew the understanding of population demographics if not taken into account. 

Malmquist and Eddington bias are not always relevant, but they are two of several selection effects astronomers need to take into account that require alternate methods.  For example, for a population study of globular clusters (dense clusters of thousands to millions of stars, all born at the same time) in another galaxy in the nearby universe, all of the clusters are effectively at the same distance from the observer's point of view, and so the volume probed is the same for all intrinsic brightnesses.  However, as the detection limit of the survey is approached, the fraction of objects detected drops.  The process for recovering the true luminosity distribution of globular clusters from the observed distribution is to perform a completeness study \citep{CompletenessWhitemore+99}.  Specifically, a large number of fake globular clusters with a distribution of brightnesses are randomly added to the observed image of the galaxy, and then the algorithm used to detect and measure the brightness of each source is run on the image which includes the false sources.  The input population of false sources is compared to the extracted population.  The fraction of detected objects as a function of input brightness is determined so that the observed distribution can be corrected for completeness (see $\S$2.3 of \citealt{Gallagher+2001} for an example).  The brightness distribution of fake sources does not have to match that of the globular clusters; it is only important that each brightness bin is well-enough sampled (has enough objects) that the uncertainties from counting statistics are small.

This type of simulation to correct for selection effects is fairly straightforward and reliable as long as there are no systematic differences in the distribution of faint versus bright sources.  For example, suppose that the most luminous sources are preferentially located in regions with high background light; in this case, the completeness correction would have to account for the negative impact of this on the detection rate.  More specifically, the completeness correction to take this effect into account would depend on more than the single parameter of observed brightness.  Not recognizing this characteristic of the true globular cluster population could lead to an underestimate of the numbers of brighter objects, and therefore result in a systematic bias.    

After decades of study, globular clusters have been well-characterized \citep{GCSReview_Harris1991}.  Population demographics are known to depend on such astrophysical properties as their ages, metallicities, and the mass and type of galaxy they inhabit. The observability of a particular globular cluster will  depend on the wavelength and sensitivity of the observation, the presence of obscuring dust in our Galaxy and the host galaxy, the projected location of the globular cluster within the host galaxy, as well as the globular cluster's brightness.  How much of these astrophysical and observational selection effects need to be accounted for in generating a completeness correction will depend on the specifics of the population under study and the characteristics of the observing program.   The background knowledge developed over decades of study of globular clusters and the other relevant aspects of astrophysics support reliable estimates of systematic biases.

As mentioned above for the specific case of globular clusters, there are selection effects induced by astrophysics, such as the effects of dust along the line of sight, that affect observations of many systems.  A screen of dust between the observer and a star will make the star's light both fainter and redder; these effects are called extinction and reddening. In the Milky Way (and other disk galaxies), dust lies preferentially in the plane of the Galaxy's disk.  It is also the case that the most luminous stars are typically in the plane, because they are from a younger population that formed there.  In the example of counting stars and binning them based on their intrinsic brightnesses, not accounting for the selection effect caused by dust that differentially affects the most luminous stars would lead one to undercount them.  The magnitude of the undercounting would also be sensitive to the color of the images, with blue images being more strongly affected.

Another astrophysical selection effect relevant in extragalactic surveys is a consequence of cosmic variance.  This refers to the non-uniformity of the distribution of extragalactic objects that can be detected if one does not sample a large-enough area.  Observations of the cosmic microwave background support taking the mass distribution of the universe to be homogeneous and isotropic to a high degree of approximation at early times.  Yet the distribution of galaxies only approaches homogeneity at very large scales.  Samples collected at smaller scales, using individual galaxies as probes, would be expected to depart from homogeneity. Evaluating samples in different small-area surveys (such as intermediate mass galaxies in the Chandra Deep Field South region, \citealt{Example_Cosmic_Variance_Ravikumar+2007}), often reveals significant differences in their distributions.

Modern galaxy surveys include wide-field imaging in many color filters, and subsequent spectroscopic follow-up.  Spectroscopy enables obtaining accurate redshifts (essential for calculating distances and therefore luminosities), and determining other galaxy properties such as star-formation rates and the ages of the dominant stellar population.  To collect sufficient light for analysis, targets for spectroscopic follow-up must be brighter than the limit of imaging surveys, and different kinds of galaxies are more amenable to spectroscopy.  For example, star-forming galaxies are typically blue and have emission lines, the latter makes measuring redshifts much easier than for quiescent (non-star-forming) galaxies (generally red) that only have absorption lines.  For absorption-line galaxies, the signal-to-noise ratio in the continuum must be higher to detect the features required to measure a redshift.  For emission-line galaxies, some redshift ranges -- including the ``redshift desert'' near $z\sim1.5$ -- have few bright emission lines in the observed-frame optical wavelength bandpass of most spectroscopic surveys.  If we consider each step of this process (measuring the light from multi-color imaging, spectroscopic target selection, and spectral analysis), there are distinct selection effects in detecting and characterizing each particular class of object.  For the DEEP2 galaxy survey, \citet{DEEP2_Newman+2013} list 7 distinct selection effects for the final sample chosen for spectroscopic follow-up: 
\begin{enumerate}
\item{Galaxy color bias due to the {\em R} magnitude limit}
\item{Loss of bright star-like objects}
\item{Misclassification of faint stars as galaxies}
\item{Loss of objects due to missing $B$ or $I$ photometry}
\item{Loss of small, distant, faint red galaxies}
\item{Loss of objects at small separations}
\item{Multiple galaxies masquerading as single galaxies}
\end{enumerate}
\par\noindent These can occur because of observing conditions, e.g., bad weather might result in missing data (item 4), the limits of the instrumentation, e.g., the spectrograph cannot observe two objects too close together (item 6), or an inability to accurately identify a galaxy based on how it presents in imaging (items 3 and 7).  This list does not even incorporate the subsequent issues that can occur once spectra are obtained, such as not finding sufficient distinguishing features to determine an accurate redshift.  Each selection effect will have a differential impact on the detection and characterization of distinct classes of galaxies.  Clearly, understanding and accounting for these interconnected selection effects rapidly becomes extremely complicated.

As galaxy surveys have become larger and more sophisticated, the tools to address selection effects have similarly developed.  Computer simulations now play an essential role, because the layers of selection effects have become too complex to account for with simple numerical corrections.  A specific technique is to use ``mock galaxy catalogs'' - a model of the true galaxy population, informed by the best understanding of galaxy demographics and evolution - and to forward-model the impact of each observational step (and its associated uncertainties) in a survey and then compare the actual observed data to the simulated observed population \citep[e.g.,][]{Coil+2007,DEEP2_Newman+2013}.  The input population in the mock catalog can be adjusted within a parameter space informed by cosmological simulations until the simulated and observed populations are consistent.  

This is a successful solution to the challenge of understanding and then correcting for observational selection effects as long as the input catalogs are a reasonable representation of the true galaxy population.  As a new survey pushes into new parameter space (e.g., by imaging at different wavelengths, pushing to fainter fluxes, or covering a larger volume), the possibility of unanticipated objects grows.  In this case, the parameter space explored in generating input catalogs can also have selection effects that generate biases; in the most extreme case, galaxies with unexpected properties -- unknown unknowns -- may simply not be included at all in the mock catalog.  

This is particularly clear in cases where it is challenging to determine a reasonable ``mock'' catalog in order to understand selection effects for objects near the detection threshold.  As an illustrative example, consider the efforts to determine accurate redshifts using photometry.  Spectroscopic determinations of redshifts are much more accurate, but cannot be feasibly used to measure the redshifts for the number of galaxies used in contemporary surveys, particularly at faint fluxes.  Astronomers have turned to easier, but coarser, photometric methods to measure redshift as an alternative.  The photometric redshift measurements are then calibrated with the spectroscopic measurements.  This requires demographic completeness of the two sets of measurements, so that they are calibrated over galaxy distributions with similar physical properties.  This is a major challenge, however, because the properties of the galaxy distributions themselves are uncertain;  it is difficult to establish how closely the catalog of galaxies based on spectroscopic observations matches that of photometric observations.  There are ongoing efforts to respond to what are called ``catastrophic failures'' of photometric redshifts (namely, cases where they depart dramatically from spectroscopic estimates), based on new types of observations and refined estimates of the systematic biases these failures induce in determinations of other parameters.  

Here it is natural to wonder whether an analog of ``experimenter's regress'' arises.\footnote{Thanks to an anonymous reviewer for raising this question.} \citet{collins1992changing} claims that there is no way to avoid circularity in identifying correct experimental results:  good results are obtained with a good experimental apparatus, and vice versa.  Anomalous results can always be rejected as the product of a malfunctioning apparatus.  According to Collins, the decision to accept certain experiments and their results is grounded in social interactions in the community and cannot be based solely on epistemic considerations.  An analogous ``observer's regress'' would regard the apparently circular trade-off between assumptions regarding the true population of astrophysical objects and selection effects.  What grounds do we have for choosing between the two, particularly for surveys extending into new parameter space at the detection threshold of existing instruments?  

Our response is similar in spirit to \citet{franklin1994avoid}'s rebuttal of Collins:  we should not amplify the legitimate challenges with calibrating experiments, or conducting astrophysical surveys, into an impossibility claim.   Frontier research faces challenging questions regarding selection effects and how to model them.  But historical cases, such as the study of globular clusters described above, reveal that the threat of circularity is only temporary:  there are several sufficiently independent lines of evidence that eventually led to a clear choice between attributing a particular result to the true population vs. a selection effect.  A culture that embraces open data policies (common practice for many observatories) also means independent teams can tackle the same datasets and apply their own suite of simulations to interpret them.  Furthermore, technological advances typically resolve some outstanding uncertainties about the nature of objects on the boundary or beyond what is currently observable; investments in developing future facilities are justified by exactly these sorts of outstanding science questions.  While this provides no guarantee that current challenges, such as that associated with redshift measurements, can be resolved in the short term, there is little support for an impossibility claim like Collins's.    

We next turn to a different kind of case, an example where the limitations of what is feasible computationally can create a novel type of selection effect.


\section{Case study: what triggers quasar activity?}

Above we described the types of knowledge, primarily regarding properties of a population of target objects and details of the observational program, that are required to handle selection effects.  These aspects of selection effects and sources of systematic bias are well-known in astrophysics, but we will now turn to a type of selection effect that has drawn less attention.  We will call this the ``problem of uncomputed alternatives'' \citep[following][]{stanford2006}:  the neglect of physically plausible scenarios that are, however, computationally intractable.  This neglect can lead to designing observational programs that have an unjustifiably narrow scope, neglecting the kind of evidence that could be relevant to assessment of the uncomputed alternative.  But by its very nature the uncomputed alternative is difficult to assess because it is computationally inaccessible:  there is at present no clear way to set up a clean comparison between observations and the alternative hypotheses.  We will illustrate this general issue through a concrete case study.

From observations in the local universe, it appears that every massive galaxy hosts a supermassive black hole at its core \citep{Kormendy+Richstone1995}.  These black holes grew primarily as quasars during the epoch known as ``cosmic noon'' ($z=1$ to 3) when the universe was approximately a quarter to a half its present age \citep{Soltan1982,Yu+Tremaine2002}.  The question of what triggers quasar activity is an area of active past and current research.  Answering that question is challenging, for reasons that intersect.  

A natural experiment that could address this question would be to observe the hosts of quasars, to characterize the galaxies they inhabit.  This is because the fuel that powers quasars comes from galaxies.  Therefore, knowing what kinds of galaxies host quasars -- for example star-forming or quiescent, with disk or spheroidal morphologies -- would put important constraints on triggering mechanisms.  However, this is more easily proposed than accomplished for several reasons.  First, a quasar often outshines the light from its host galaxy by factors of up to 1000.  Second, quasar host galaxies at the distances commensurate with cosmic noon have small angular extents, on the order of $\sim1^{\prime\prime}$.  From the ground, this is close to the angular resolution of most telescopes (from the smearing of the atmosphere), and thus separating the lower surface brightness host galaxy from the very bright quasar in its center in an image is typically not possible.  This challenge of ground-based observations is why studying quasar host galaxies has been a science focus for both the Hubble and Webb Space Telescopes. 

The first samples with Hubble imaging of quasar host galaxies showed a range of morphologies, including some indicative of interacting galaxies \citep[e.g.,][]{Hutchings+Morris1995,Bahcall+1997}.  The varied selection criteria and relatively small sample sizes (a few to 20 objects) make drawing conclusions from the fraction of observed galaxies that showed evidence of mergers challenging.  For example, some of the galaxies chosen for Hubble imaging were selected based on evidence from ground-based observations for extended, asymmetric structures, and so it is not surprising that these galaxies were often found to be likely merger remnants \citep{Hutchings+1994}.  Time on a valuable resource such as Hubble is allocated through a very competitive process, and an observing program that is more likely to yield a positive result (such as a clear detection of interesting structure) is more likely to get chosen.\footnote{This illustrates another potential selection effect in astronomy, that of the telescope time allocation committee.}

A second empirical path would be to look at nearby quasars, where these observational challenges can be mitigated because the host galaxies are significantly larger and have higher surface brightnesses.  Locally, many luminous quasars are found in `warm' ultra-luminous infrared galaxies, the highest luminosity galaxies (with $L_{\rm IR} \ge 10^{12}$~M$_{\odot}$), with infrared properties that indicate higher temperature dust, most plausibly heated by a quasar (as opposed to active star formation) \citep{Sanders+1988}.  High-resolution Hubble Space Telescope imaging of some of these galaxies revealed signatures of recent galaxy mergers, including tidal features and young, massive star clusters whose formation could be triggered by a merger event \citep{Surace+1998}.  Mergers of gas-rich disk galaxies are plausible triggers for quasar activity, as the collision of gas clouds can efficiently shed sufficient angular momentum to drive gas towards the gravitational center of the merger remnant, where the supermassive black hole is found.  

With the first generation of galaxy-merger simulations that included gas (which can dissipate energy and cool radiatively) and stars (which behave as collisionless particles that only interact gravitationally), the theoretical support for the idea that quasars could be caused by mergers was demonstrated \citep{Barnes+Hernquist1991}.  However, it should be recognized that the models themselves did not include supermassive black holes, nor did they have sufficient spatial resolution to follow the gas to the center of the potential well at the scales of the gravitational sphere of influence of a supermassive black hole.

One of the challenges of setting up a galaxy-merger simulation is choosing appropriate initial conditions for the encounter from among a very large parameter space of possibilities.  For example, the relative initial positions and velocities for each galaxy, the inclinations of the disks, and the sense of their motions (clockwise or counterclockwise), all impact on the progress of the merger and the final outcome.  Furthermore, the structure of the galaxy itself (such as how prominent the central bulge is relative to the disk) impacts the gas flows within the galaxies in the course of merging and thus the amount and timing of induced star formation \citep{Mihos+Hernquist1996}.  Since the first galaxy-merger computer simulations of \citet{Toomre+Toomre1972}, the touchstones for these merger simulations are often local ultraluminous infrared galaxies, and one measure of success claimed by simulators is to match (at some point in the progression of a merger) well-known examples of merging pairs.  Full-blown merger simulations are computationally expensive, and so judicious choices of initial conditions are important.  \citet{Toomre+Toomre1972} chose parabolic passages, and were able to come up with reasonable representations of four well-known merging galaxy pairs.  

Taken together, both the simulations of galaxy mergers and observations of nearby quasar host galaxies provided a consistent picture whereby a merger of two gas-rich disk galaxies could drive gas towards the center of the potential well of the merger remnant and provide the fuel to power a quasar and grow a supermassive black hole.  Empirically, this scenario holds up well in the local universe, where both mergers of gas-rich galaxies and quasars are quite rare.  At earlier times, galaxies were more numerous and closer together, and quasars were both more common and more luminous.  So, does this story, well-supported at low redshift, also hold at $z\sim2$?

The observational story at higher redshift is complicated, because it is still challenging to separate out the light from quasar host galaxies.  Signatures of mergers such as tidal tails and young massive star clusters become significantly harder to resolve spatially.  In this case, the role of simulations becomes even more important.  From the first generation simulations of \citet{Barnes+Hernquist1991}, subsequent researchers made correspondingly more sophisticated merger simulations \citep[e.g.,][]{DiMatteo+2005,Hopkins+2005}, that supported the original success of mergers accounting for quasar activity at early times.  Typically, the initial conditions for galaxy interactions are generated from low-resolution cosmological simulations, and then a higher-resolution simulation is performed to follow the subsequent evolution, with analytic prescriptions for the onset of star formation and black-hole feeding that are below the spatial resolution of the galaxy simulations.  

The case for inferring that what happens locally also works at higher redshifts breaks down when we consider the significant evolution of galaxies over billions of years.  In particular, at higher redshifts, disk galaxies have a higher fraction of their baryonic mass in gas, and also have dynamically `hotter' disks, with significant vertical (in addition to primarily rotational) motions.  A consequence of these structural properties is that star-forming regions are typically larger because it takes more mass to cause gravitational collapse against gas motions \citep{LargeClumps_Elmegreen+2007}.  Next, gravitational instabilities in the disk gas, such as spiral arms and bars, can happen through secular evolution, without an external trigger.\footnote{Spatially resolved kinematics of a small sample of star-forming galaxies at $z\sim1.5$ indicate galaxies with both disk-like and merger-remnant structures, and higher velocity dispersions in star-forming clumps than typically seen at low $z$ \citep{HighzDisks_Mieda+2016}.}  Bars are evidence of radial motions in the gas, and are effective at funneling gas to smaller radii.  These factors together mean that a starburst episode coupled (or followed by) quasar activity can plausibly happen without significant dynamical shocks triggered by a merger \citep{Secular_Hopkins+2010}.  

In addition, quasars at $z\sim1$ are found typically in galaxy group environments \citep{Coil+2007}, with several galaxies gravitationally bound to each other.  With a handful of galaxies (rather than just two), gravitational interactions become much more complex, and are less likely to lead to a merger of a pair.  However, in groups galaxies do interact gravitationally, but the effects -- such as low surface brightness tidal features and depletion of cold gas reserves -- can be much more subtle than the dramatic impacts of a merger \citep[e.g.,][]{HCG7_Konstantopoulos+2010}.  These empirical results on secular disk evolution and small galaxy group interactions suggest alternate pathways for triggering quasar activity accompanied by active star formation than the merger of a gas-rich galaxy pair.  Such subtle effects would also be challenging to detect beyond the local universe.  

One reason that the pair-merger pathway to quasar activity has been so widely accepted is the success of the computer simulations of the physical system.  Observations of any single system will necessarily capture only a moment in time, and a collection of observations of different systems has to be put together into a coherent picture to understand evolution over billions of years.  Computer simulations thus serve an essential role in filling in the time gaps, and following a single type of system over time.\footnote{This is a further instance of a role for simulations that \citet{jacquart2020observations} emphasizes, namely amplifying astrophysical observations --- in this case, moving from isolated instants to an evolutionary trajectory for a type of system.}   As a recent example, \citet{PairSimulations_Moreno+2021} investigate the effects of galaxy-pair interactions on star formation within each galaxy (black-hole fueling is not included in the simulations) with a suite of 24 galaxy-pair simulations (varying the initial conditions).

One should also consider, however, which computer simulations are {\em not} being done.  A specific example is a simulation of a small group of galaxies to investigate if (and how) modest and perhaps recurrent gravitational interactions between more than one galaxy could trigger star formation and quasar activity. Practical constraints explain the lack of simulations of this type of system to address the question. First,  a single simulation of even three galaxies would be computationally very challenging.  Second, such a simulation would also require choosing initial conditions (such as galaxy properties and relative positions and velocities) from a very large parameter space of potential values.  Running a large number of simulations to investigate the influence of initial conditions would be computationally extremely expensive.  But there are no physical grounds to rule out this kind of interaction.  This is an example of an ``uncomputed alternative,'' a reasonable hypothesis that has not been explored because the simulations required are not currently feasible.       

One of the plausible explanations for the triggering of quasar activity has thus not been explored using simulations, and therefore is not subject to detailed observational evaluation.  This is an example of a novel type of selection effect induced by what is computationally tractable that is limiting the space of hypotheses under consideration.

There are several consequences of a computational selection effect.  One is a limitation on the types of observational programs that may be undertaken to test the merger-trigger hypothesis, and also how those data are interpreted.  For example, the empirical study of \citet{Pairs_AGN_Ellison+2011} considered low-redshift pairs of galaxies to see if evidence for accretion onto a black hole (spectroscopic identification as an active galactic nucleus) was correlated with being classified as a close-separation pair.  Though higher multiples (triples or more) were not selected against, the target sample and control sample of isolated galaxies were all chosen from the Sloan Digital Sky Survey galaxy sample, which has a relatively bright flux limit and (because of instrumental limitations) a known high level of spectral incompleteness for close galaxy separations.  The authors are well aware of the potential consequences of these selection effects, but limit their discussion to the evaluation of pair-wise interactions, described as merger candidates, versus the alternate pathway of secular disk evolution to explain black-hole fueling.

In another empirical study, \citet{SFR_Pairs_Patton+2013} considered observed enhancements in star formation in galaxy pairs and used a suite of 75 merger simulations for comparison.  Though their discussion of sample selection of the target sample and control sample of isolated galaxies accounted for local environment (acknowledging that most galaxies are found in groups and clusters), the simulations themselves did not incorporate more than two galaxies.  In this case, the use of simulations to reveal the mechanism for the observed increased star formation rate of paired galaxies provides less convincing evidence.

The impact of this ``problem of uncomputed alternatives'' resembles that of \citet{stanford2006}'s problem of unconceived alternatives:  the force of an eliminative argument in favor of a hypothesis depends on whether all reasonable alternatives have been considered.  In our view, the example above illustrates a viable physical mechanism for triggering quasar activity that has not been eliminated, and the case in favor of the predominance of the pair-merger pathway is hence less compelling.  (That is not to downplay the importance of the positive case in favor of this proposal:  it is based on extrapolating a successful account from low redshift back to the earlier universe.  But it does call into question the epistemic support added by the simulation studies.)  

There are also two contrasts with Stanford's account worth noting.  The assessment of the space of ``plausible'' competing hypotheses is challenging, and Stanford's historical arguments are intended to illustrate ways in which scientists have routinely failed to consider viable alternatives in the form of radically different theories. This example has a different character: the apparent success of simulations of (relatively speaking) simple cases may lead to an overconfidence in extrapolating to more complex cases, where other causal factors may be in play.  In the case we discuss, the alternatives involve different assessments of what physical interactions are relevant to a particular phenomenon, but do not raise questions about the underlying physical theories.  The failing is not insufficient exploration of the space of possible theories, but insufficient exploration of how to treat complex situations with existing theory.

But the second contrast is more significant.  The failure to include ``uncomputed alternatives'' undermines an eliminative argument, but it also has a more subtle impact on the interpretation of observations.  Analyzing potential selection effects requires a comprehensive understanding of how the properties of a target population interact with the observational program, and any biases that these produce.  It is much harder to characterize the impact of the observing programs that are not undertaken because of how the science question is formulated.  Specifically, an observing program addressing the question of whether group interactions (without mergers) can trigger quasars at cosmic noon would be fundamentally different than the programs of \citet{Pairs_AGN_Ellison+2011} and \citet{SFR_Pairs_Patton+2013} described above.
\
\section{Conclusion}

Astronomers use simulations routinely in order to model selection effects.  In cases like large galaxy surveys, inter-related selection effects from a variety of sources, such as details of the instrument and observational program to the astrophysics of the target systems, can no longer be treated through individual numerical corrections.  As we have described above, astronomers instead simulate the expected output of an observing program for a ``mock'' catalog of sources, and use the comparison of these extracted results to actual observations to assess and account for selection effects.  We have described a few concrete examples above, with the aim of illustrating this technique in more detail and clarifying the kinds of background knowledge that are needed for it to be reliable.  Establishing reliability is particularly difficult when uncertainties regarding selection effects are compounded with uncertainties regarding the population of target objects.  Finally, we identified a novel kind of computational selection effect that we called the ``problem of uncomputed alternatives.''  In some cases, physically reasonable proposals simply cannot be followed through computationally, at least at present, to determine their observational signatures.  The neglect of these possibilities may lead to the design of observational programs that cannot reveal problems with simpler, albeit incomplete or incorrect, alternative hypotheses.      

\bibliographystyle{psalike}
\bibliography{psa2018.bib}

\end{document}